# Magnonics: Experiment to Prove the Concept

V.V. Kruglyak and R.J. Hicken

School of Physics, University of Exeter, Stocker Road, Exeter, EX4 4QL, UK

An experimental scheme for studying spin wave propagation across thin film samples is proposed. An experiment upon a periodically layered nanowire is numerically simulated, while the sample might equally well be a continuous film or an array of elements (e.g. nanowires) that either have uniform composition or are periodically layered as in a magnonic crystal. The experiments could be extended to study domain wall induced spin wave phase shifts, and used for creation of the spin wave magnetic logic devices.

Following recent developments in the experimental techniques applicable to magnetic systems, the idea of magnonic devices, in which spin waves (magnons) could be used for transmitting and processing information, has received a increased attention in the magnetics community[1-21]. Standing spin waves have been observed in micron size thin film metallic elements by means of the time resolved scanning Kerr microscopy (TRSKM) and Brillouin light scattering (BLS)[9,10,12,17]. The propagation of spin waves within much larger ferrite films, that are either continuous or periodically structured in the film plane, has been investigated using the space and time resolved BLS and inductive detection techniques[1-8,13]. It has been proposed that spin waves can be used to investigate domain wall properties within an Aharonov-Bohm geometry[14], and in magnetic logic devices[19]. Typically, the spin waves considered in those works had wavelengths of several micrometers and frequencies within the sub-ten Gigahertz range, and were of magneto-dipole rather than exchange character. At the same time, exchange spin waves have wavelengths of just a few nanometers and frequencies of tens of Gigahertz, and hence could be used in much faster devices of nanometer size. While exchange spin waves propagating perpendicular to the surface of thick ferrite films have been studied in the frequency domain in Ref. , a time domain experiment is required to take advantage of effects such as those predicted in Ref. 14.

In this paper, we propose an all-optical time domain experiment for investigation of propagating exchange spin waves in nanoscale metallic samples. We describe how spin waves with wavelengths of down to ten nanometers may be excited and to detected, and how the frequency of the excitation may be to controlled. We use the Object Oriented

Micromagnetic Framework (OOMMF)[22] to model the signal resulting from the spin wave propagation along a periodically layered nanowire[16,18].

The principle of the proposed experiment is shown in Figure 1. Spin waves are excited on the front surface of the sample, propagate across its thickness, and are detected on the rear surface. The excitation of short wavelength spin waves requires magnetic field pulses localized within a nanoscale region of the sample. For this purpose, an antiferromagnetic exchange bias layer is deposited onto the front surface of the sample. When irradiated by an intense femtosecond laser pulse, this layer exerts a pulsed effective magnetic field upon the adjacent ferromagnet[23]. The newly demonstrated optical excitation of an ultrafast antiferromagnetic to ferromagnetic phase transition in exchange spring materials represents another way of generating short localized effective field pulses[24]. In each case, the effective field pulses originate from the exchange interaction and hence are localized in the few monolayer thick interface region of the sample. To study propagating rather then standing spin waves, the sample should be thick compared to the depth of localization of the transient effective field, although thin compared to the spin wave decay length so that the signal detected on the rear surface is of measurable amplitude. The sample may be either a continuous film or an array of elements (e.g. nanowires), that has either uniform composition or is inhomogeneous (e.g. periodically layered as in the case of a magnonic crystal[18]). TRSKM or time resolved BLS could be used to detect the transmitted spin waves. It is essential that the sample is metallic so that the optical field is confined to the surface region. Then the magnetic response is sampled within the optical skin depth, i.e. about ten nanometers, which provides the lower limit for the spin wave wavelength detectable with this technique. The wavelength sensitivity can be further enhanced, virtually down to the atomic length, if magnetic second harmonic generation (MSHG) is used for detection.

A single optical pump pulse may excite many modes with frequencies corresponding to the Fourier spectrum of the optically generated effective field pulse. For some applications, spin waves with a certain frequency may be preferred. As shown in Figure 2, a combination of a Fabry-Perot etalon like assembly and a filter of varying optical density can be used to form an optical pulse train. The latter leads to an effective field pulse with a Fourier spectrum that contains a strong peak at a frequency corresponding to the time delay between subsequent optical pump pulses.

To explore what kind of information can be obtained from the proposed experiment, we have considered a 1 µm thick film representing an array of periodically layered nanowires embedded into a nonmagnetic matrix with an exchange bias layer deposited on one surface. The nanowires lie perpendicular to the film surface and have a square cross-section with 20 nm side. Each nanowire consists of two different alternating homogeneous ferromagnetic layers with thickness of 20 nm, and hence has a period of 40 nm. The two layers have an exchange parameter of 13 $\frac{\text{pJ}}{\text{m}}$, a uniaxial anisotropy of 0.5 $\frac{\text{kJ}}{\text{m}^3}$, a Gilbert damping parameter of 0.01, and a $g$-factor of 2.1, while they have different saturation magnetization values of 1 and 0.9 $\frac{\text{MA}}{\text{m}}$. The external bias magnetic field of 0.01 T and the easy anisotropy axis are aligned parallel to the axis of the nanowire. The interfaces are assumed to be sharp and flat, and lie perpendicular to the easy axis. The nanowires are assumed to be sufficiently separated that the inter-wire dipolar interaction can be neglected, and so we have performed numerical simulations for an isolated nanowire only. The exchange bias layer is assumed to have identical magnetic properties as that studied in Ref. 23. Therefore we have assumed that, upon excitation by a picosecond laser pulse, a pulsed exchange bias field of 1 mT amplitude, with rise time of 20 ps and decay time of 170 ps, is created in the 2 nm thick end region of the nanowire adjacent to the exchange bias layer. The simulations were performed with cells of 0.5 x 20 x 20 $nm^3$ size, and the magnetic state of the nanowire was recorded every 5 ps within 5 ns after the excitation.

The results of simulations are presented in Figure 3. On the "time-distance" plot, the propagation of the perturbation at the left end of the wire is first seen in the form of almost straight beams of propagating spin waves slightly distorted due to the interference with reflections from interlayer interfaces. The shortest wavelength modes reach the right end in less than 200 ps, while the main portion of spin waves arrives after about 700 ps. After that a characteristic "grating" pattern of interference between the forward modes and those reflected from the right wire end is formed. A mode of smaller frequency as compared to that of the propagating spin waves is confined near the wire end by the nonuniform demagnetizing field[25], and serves as a source of propagating modes. Figure 3 (b) shows a time resolved signal obtained by averaging the response of the last layer of the nanowire, i.e. similar to the signal measured in the proposed experiment if the skin depth of the probe is about 20 nm.

Figure 4 (a) shows a two dimensional fast Fourier transform of the "time-distance" data array, a portion of which is presented in Figure 3 (a). One can immediately identify the parabolic dispersion relation of the exchange dominated spin waves. Branching is observed at the points where the inverse wavelength is equal to (12.5 $i$) $\mu m^{-1}$ ($i$ is integer), i.e. at the Brillouin zone boundaries[18]. Figure 4 (b) shows Fourier spectra calculated from time resolved signals obtained from spatial averages of different portions of the nanowire. Within the continuous spectrum of propagating spin waves, one can see clear dips that correspond to the band gaps at the Brillouin zone boundaries[18,26]. We note that the spectra from the last layer and the last cell (of the simulation) of the wire are similar to those which could be observed experimentally using TRSKM and MSHG, respectively.

In summary, we have proposed an experimental scheme for studying spin wave propagation across thin film samples. We have described how short wavelengths spin waves may be excited and detected, and how the frequency of excited spin waves may be controlled within the Gigahertz regime. While micromagnetic simulations have been presented for experiments performed upon a periodically layered nanowire, the sample might equally well be a continuous film or an array of elements (e.g. nanowires) that either have uniform composition or are periodically layered as in a magnonic crystal. The experiments could be extended to study domain wall induced phase shifts within the Aharonov-Bohm geometry proposed in Ref. 14, and constitute the basis for realization of the spin wave magnetic logic devices proposed in Ref. 19.

The authors acknowledge the financial support of the UK Engineering and Physical Sciences Research Council (EPSRC) and the New Energy and Industrial Technology Development Organization (NEDO).

List of figure captions

Figure 1	(Color online) A schematic of the proposed experiment is shown. The exchange bias/spring layer on the right hand side is irradiated by the pump laser pulse, and hence a pulsed torque is exerted upon the adjacent thin film multilayer sample, which also may be an array of uniform or multilayered elements. The torque excites spin waves that propagate towards the left hand side surface of the sample where they are detected optically.

Figure 2	(a) An exchange bias field pulse is shown that was obtained by a superposition of five pulses with rise time of 20 ps and decay time of 170 ps each. The subsequent pulses were delayed by 50 ps. (b) The Fourier spectrum of the pulse in (a) is shown. The insets shows the Fabry-Perot like assembly used to form the optical pump pulse train leading to a superposition of three exchange bias field pulses. (1) is a mirror; (2) is a 10/90 beam splitter; (3) is a gradient neutral density filter; (4) is a lens; (5) is the sample.

Figure 3  (a) The results of the simulations are shown on the "time-distance" plane. The grey scale represents the magnetization component perpendicular both to the nanowire axis and the pulsed effective field. (b) The time resolved signal obtained by averaging the response of the last (white) layer of the nanowire is shown.

Figure 4  (a) The two dimensional fast Fourier transform of the "time-distance" data array from Figure 3 (a) is shown. (b) Fourier spectra calculated from average time resolved signals from the last cell (of the OOMMF simulations), the last layer, and the first layer of the nanowire are presented in the top, middle, and bottom panels, respectively. The arrows mark positions of the band gaps.

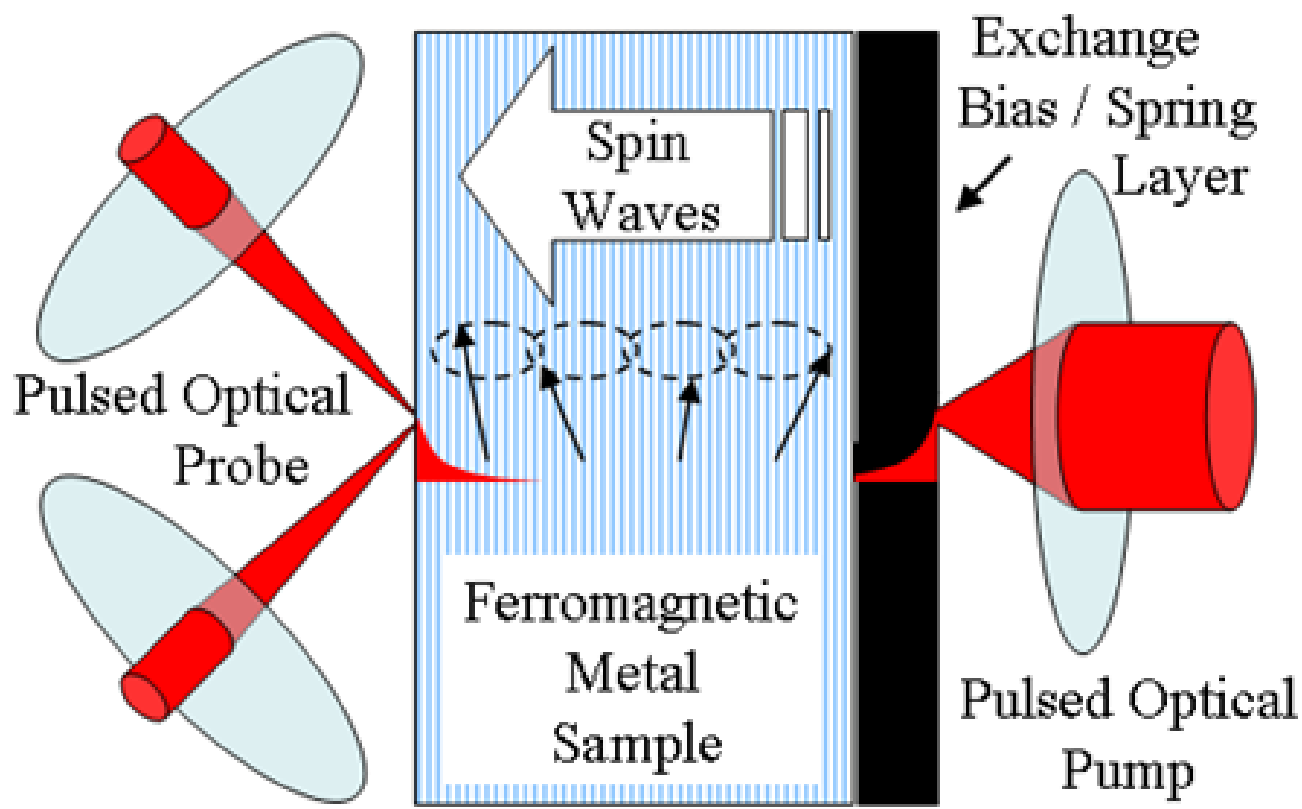

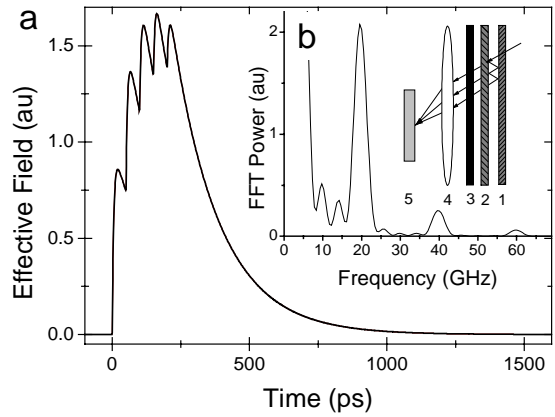

Figure 2 (a) An exchange bias field pulse is shown that was obtained by a superposition of five pulses with rise time of 20 ps and decay time of 170 ps each. The subsequent pulses were delayed by 50 ps. (b) The Fourier spectrum of the pulse in (a) is shown. The insets shows the Fabry-Perot like assembly used to form the optical pump pulse train leading to a superposition of three exchange bias field pulses. (1) is a mirror; (2) is a 10/90 beam splitter; (3) is a gradient neutral density filter; (4) is a lens; (5) is the sample.

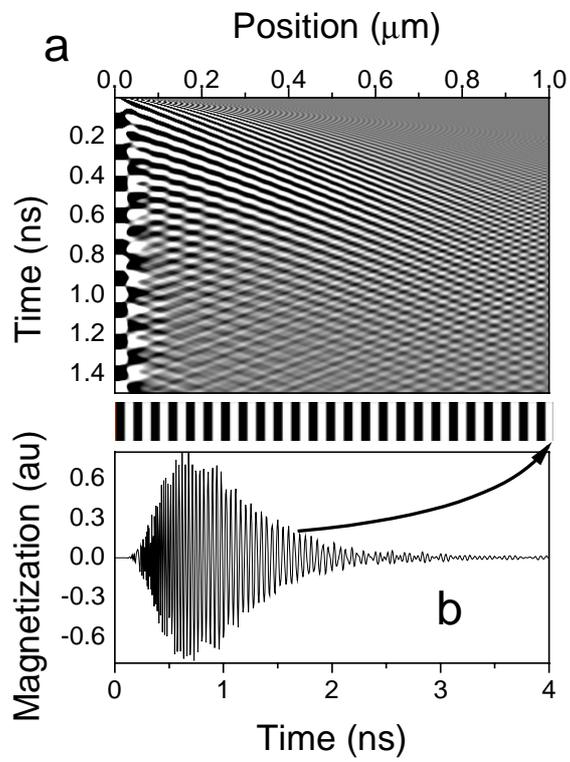

Figure 3 (a) The results of the simulations are shown on the "time-distance" plane. The grey scale represents the magnetization component perpendicular both to the nanowire axis and the pulsed effective field. (b) The time resolved signal obtained by averaging the response of the last (white) layer of the nanowire is shown.

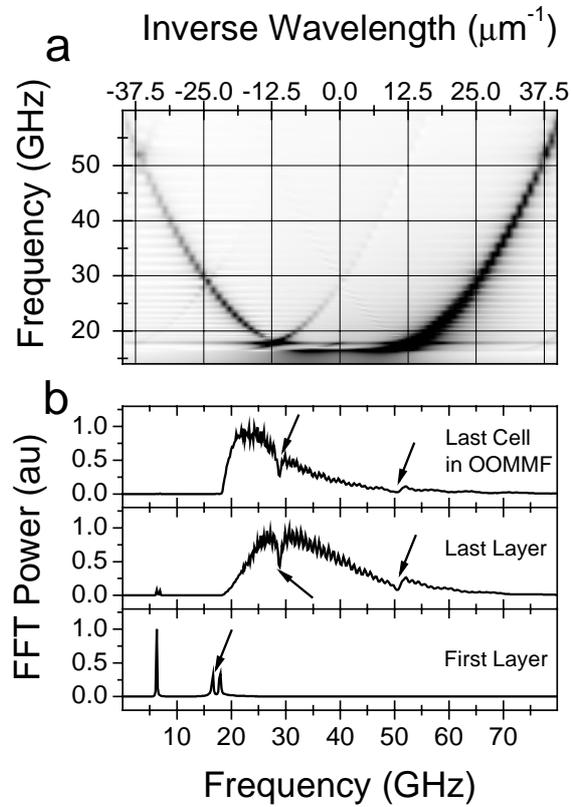

Figure 4 (a) The two dimensional fast Fourier transform of the "time-distance" data array from Figure 3 (a) is shown. (b) Fourier spectra calculated from average time resolved signals from the last cell (of the OOMMF simulations), the last layer, and the first layer of the nanowire are presented in the top, middle, and bottom panels, respectively. The arrows mark positions of the band gaps.